\documentclass[twocolumn,english,amsmath,amssymb,aps,prl,showpacs]{revtex4}
\usepackage[T1]{fontenc}
\usepackage[latin9]{inputenc}
\usepackage{graphicx}

\makeatletter

\providecommand{\tabularnewline}{\\}

\@ifundefined{textcolor}{}
{%
 \definecolor{BLACK}{gray}{0}
 \definecolor{WHITE}{gray}{1}
 \definecolor{RED}{rgb}{1,0,0}
 \definecolor{GREEN}{rgb}{0,1,0}
 \definecolor{BLUE}{rgb}{0,0,1}
 \definecolor{CYAN}{cmyk}{1,0,0,0}
 \definecolor{MAGENTA}{cmyk}{0,1,0,0}
 \definecolor{YELLOW}{cmyk}{0,0,1,0}
 }

\@ifundefined{definecolor}
 {\@ifundefined{definecolor}
 {\@ifundefined{definecolor}
 {\usepackage{color}}{}
}{}
}{}

\newcommand{\bra}[1]{\langle #1 |}
\newcommand{\ket}[1]{| #1 \rangle}

\newcommand{\rdm}[3]{\bra{#1}| #2 | \ket{#3}}

\makeatother

\usepackage{babel}

\makeatother

\usepackage{babel}

\makeatother

\usepackage{babel}

\begin{document}

\title{Optically induced spin polarisation of the $\mbox{NV}^{-}$ centre
in diamond: role of electron-vibration interaction}

\author{N.B. Manson$^{1}$, L. Rogers$^{1}$, M.W. Doherty$^{2}$, L.C.L.
Hollenberg$^{2}$}

\affiliation{$^{1}$Laser Physics Centre, Research School of Physics and Engineering,
Australian National University, Australian Capital Territory, Australia\\
 $^{2}$School of Physics, University of Melbourne, Victoria, Australia}

\date{5-11-2010}
\begin{abstract}
The novel aspect of the centre ($\mbox{NV}^{-}$) in diamond is the
high degree of spin polarisation achieved through optical illumination.
In this paper it is shown that the spin polarisation occurs as a consequence
of an electron-vibration interaction combined with spin-orbit interaction,
and an electronic model involving these interactions is developed
to account for the observed polarisation. 
\end{abstract}

\pacs{76.70.Hb; 71.70.Fk; 71.70.Ej; 76.30.Mi}

\maketitle

\section{Introduction}

Optically induced spin polarisation of the negatively charged nitrogen-vacancy
centre in diamond ($\mbox{NV}^{-}$) has been known for a considerable
time but there has not been a satisfactory account of how it arises.
This lack of explanation is of concern considering that spin polarisation
is the key property that sets this centre apart from all other optically
active centres in solids and has enabled diamonds to be used for many
new exciting applications such as magnetic sensing \cite{chernobrod2005_spin,degen2008_scanning,taylor2008_high-sensitivity,maze2008_nanoscale,balasubramanian2008_nanoscale,hall2009_sensing,cole2009_scanning,hall2010_monitoring},
probing biological materials \cite{fu2007_characterization,chang2008_mass,tisler2009_fluorescence},
and quantum information processing \cite{gaebel2006_room-temperature,dutt2007_quantum,togan2010_quantum,neumann2010_quantum,fuchs2010_excited-state}.
In this work a transition within the spin-polarising decay path is
studied using uniaxial stress. The order of the intermediate states
involved is established, but more significantly it is shown that there
is electron-vibration interaction associated with the lower level.
This interaction is also involved in the important decay to the ground
state, and it is clear that electron-vibration interaction is the
factor that has been overlooked in previous treatments of the optical
pumping cycle. Electron-vibration interaction plays a vital role in
giving rise to spin polarisation and by including it in a model of
the centre we can account for the observed level of spin polarisation.

\section{Uniaxial stress}

\begin{figure}
\includegraphics{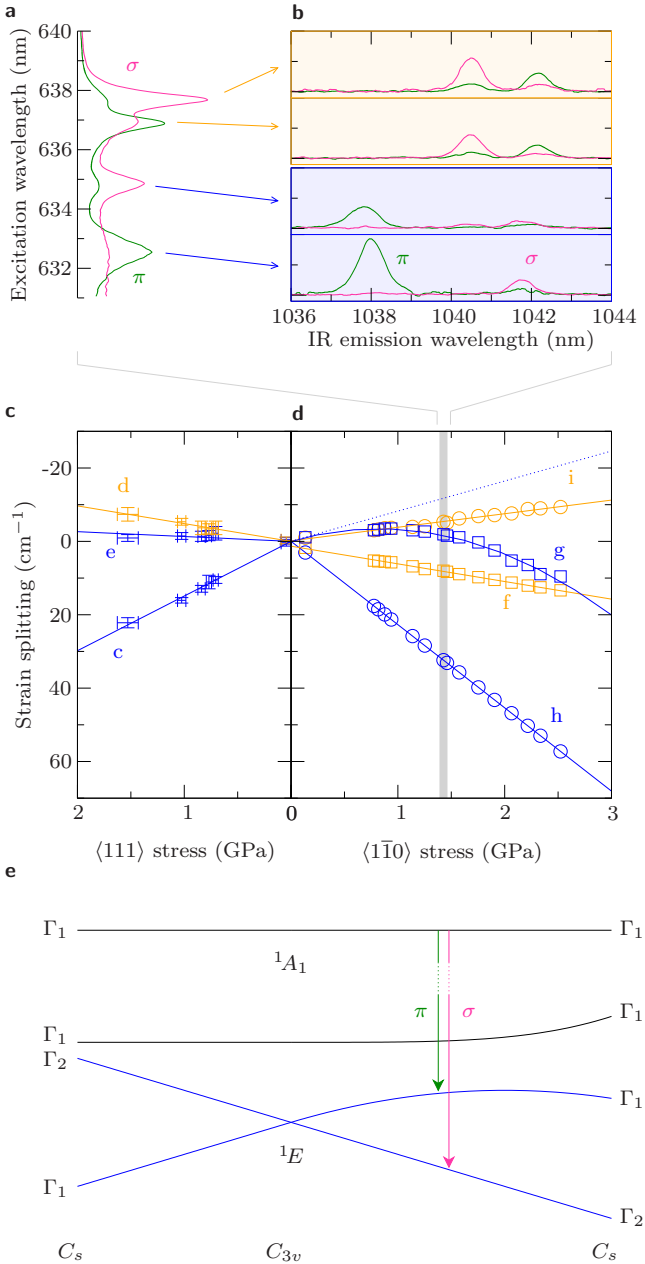}

\caption{\textbf{ZPL splittings due to uniaxial stress. }\textsf{\textbf{a}},
The 637 nm visible ZPL measured in excitation. \textsf{\textbf{b}},
Corresponding IR photoluminescence spectra for resonant excitation
on each visible peak. Resonantly exciting allowed selection of the
multiple defect orientations. \textsf{\textbf{c}}, Strain splitting
for the IR line for $\langle111\rangle$ stress. \textsf{\textbf{d}},
Strain splitting for the IR line for $\langle110\rangle$ stress.
\textsf{\textbf{e}}, Schematic diagram of the splitting and mixing
levels involved in the IR transition. \label{fig:strain-measurements}}

\end{figure}

The $\mbox{NV}^{-}$ centre in diamond has trigonal symmetry ($C_{3v}$)
and a zero-phonon line (ZPL) at $637\,\mbox{nm}$ ($1.945\,\mbox{eV}$)
corresponding to a $^{3}\! A_{2}-^{3}\! E$ transition that can be
observed in absorption and emission \cite{davies1976_optical}. An
additional ZPL at $1042.6\,\mbox{nm}$ ($1.19\,\mbox{eV}$) has been
observed for a $^{1}\! A_{1}-^{1}\! E$ transition in emission \cite{rogers2008_infrared}.
These lines and their inter-relationship are studied in this work
using uniaxial stress. The line at $637\,\mbox{nm}$ is measured in
excitation is split by applied uniaxial stress as shown in figure
\ref{fig:strain-measurements}a. The magnitude of the splittings are
in good correspondence with those obtained in absorption by Hamer
and Davies \cite{davies1976_optical}. When the laser is tuned to
one of the peaks in the spectrum, only one set of equivalent $\mbox{NV}^{-}$
orientations is excited and the infra red spectrum can be measured
for the same set of orientations as in figure \ref{fig:strain-measurements}b.
As assignments for the visible features are already known, this selective
excitation technique allows the infrared lines to be reliably assigned
(one modification is required from that given previously \cite{rogers2008_infrared}).The
variation of line positions of the infra red transition can be more
conveniently obtained by exciting using an intense green beam where
all orientations are excited simultaneously. The variation with stress
is shown in figure \ref{fig:strain-measurements}c and \ref{fig:strain-measurements}d.

\begin{figure}
\includegraphics{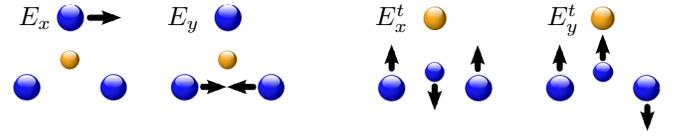}

\caption{\textbf{Symmetry adapted distortions.} $E_{x}$ and $E_{y}$ are in
the plane of the carbon atoms, while $E_{x}^{t}$ and $E_{y}^{t}$
are out of the plane. \label{fig:Symmetry-adapted-distortions.}}

\end{figure}

The zero-phonon line splittings are nearly all linear, and the strength
of the various interactions giving rise to the shifts and splittings
can be determined. The ZPL corresponds to $A-E$ transitions at the
trigonal sites, and the interaction for this situation has been treated
previously and can be expressed as \begin{eqnarray}
H_{s} & = & A_{1}(s_{xx}+s_{yy}+s_{zz})+A_{1}^{t}(s_{yz}+s_{zx}+s_{xy})\notag\\
 & + & E_{X}(s_{xx}+s_{yy}-2s_{zz})+E_{Y}\sqrt{3}(s_{xx}-s_{yy})\notag\\
 & + & E_{X}^{t}(s_{yz}+s_{zx}-2s_{xy})+E_{Y}^{t}\sqrt{3}(s_{yz}-s_{zx})\label{eq:stresshamiltonian}\end{eqnarray}
where $A_{1}$ and $E_{X}$, $E_{Y}$, $E_{X}^{t}$, $E_{Y}^{t}$
are electronic operators and $s{}_{ij}$ are components of the stress
tensor \cite{hughes1967_uniaxial,davies1976_optical}. The notation
is that adopted by \cite{davies1976_optical} except a suffix $t$
is used rather than a dash to indicate the terms which arise from
$T_{2}$ terms in $T_{d}$ symmetry. The $E_{X}$, $E_{Y}$ symmetry-related
distortions are in the plane of the three carbons, whereas those associated
with $E_{X}^{t}$, $E_{Y}^{t}$ are out of the plane (figure \ref{fig:Symmetry-adapted-distortions.}).
The shift of levels for the various stress directions have been given
by \cite{davies1976_optical} using parameters $A1$, $A2$, $B$
and $C$ associated with the interaction terms $A_{1}$, $A_{1}^{t}$,
$E$ and $E^{t}$ respectively. These splitting parameters for the
infrared transitions are summarised in table \ref{tab:stressparameters}.
The visible $\mbox{NV}^{-}$transition at $637\,\mbox{nm}$ and the
$\mbox{NV}^{0}$transition at $575\,\mbox{nm}$ also correspond to
$A-E$ transitions and the values for these transitions are given
for comparison.

\begin{figure}
\includegraphics{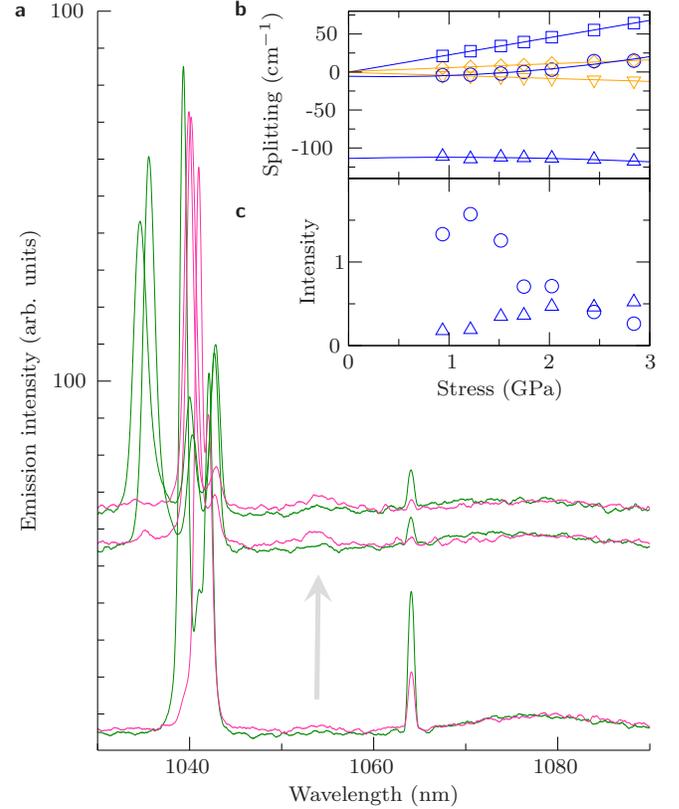}

\caption{\textbf{New spectral line. }\textsf{\textbf{a}}, Emission spectra
in $\pi$ and $\sigma$ polarisations for increasing applied stress,
displaced vertically for clarity. The sharp feature at $1064\,\mbox{nm}$
is due to laser scatter, and between it an the ZPL a broader weak
feature is present at higher strains as indicated by the arrow. \textsf{\textbf{b}},
The splitting pattern from Figure \ref{fig:strain-measurements}d
shown with the new line that appears at $-115\,\mbox{cm}^{-1}$. \textsf{\textbf{c}},
The new line gains intensity at the expense of the nonlinear ZPL component.\label{fig:New-spectral-line.}}

\end{figure}

There is one transition where the displacement with stress is not
linear. This is the case for $[110]$ stress where a line curves to
higher energy with increasing stress. For the same situation a new
line is observed at $-115\,\mbox{cm}{}^{-1}$ as shown in figure \ref{fig:New-spectral-line.}.
The new line is displaced in the reverse sense and gains intensity
at the expense of the former, indicating that the two levels are interacting.
The line is on the low energy side of the emission spectrum, and so
corresponds to a level either below the emitting level or above the
lower level. The first option would lead to strong emission from the
new level particularly once the transition becomes allowed, which
is not observed, and so the new level must be on the high energy side
of the lower state. It only gains intensity as a consequence of mixing
with one of the stress split components of the lower state, which
must be the orbital $E$ to split. This provides conclusive evidence
that the $^{1}\! E$ is the lower level and the $^{1}\! A_{1}$ the
emitting level, resolving the recent contention surrounding the order
of the levels involved in the infrared transition \cite{manson2007_issues,rogers2008_infrared,gali2008_ab,delaney2010_spin-polarization}. 

For a $[110]$ stress orientation, a reflection plane is maintained
resulting in $C{}_{s}$ symmetry. A $\Gamma_{1}(E_{x})$ component
of $^{1}\! E$ is shifted up in energy and as there is interaction,
the new state must likewise have $\Gamma_{1}$ symmetry. As there
is no splitting the level will have $A_{1}$ symmetry in $C_{3v}$.
There is no interaction when the stress is applied along the $[001]$
direction as in this case a $\Gamma_{2}\,(E_{Y})$ state is increased
in energy and will not interact with a $\Gamma_{1}$ (illustrated
in figure \ref{fig:strain-measurements}e). Note the reverse situation
occurs for the $\mbox{NV}{}^{0}$ centre, with a new line and non-linear
shifts for $[111]$ (and $[001]$) stress being reported by Davies
\cite{davies1979_dynamic}. In that centre a linear shift is observed
for a $[110]$ stress and this is consistent with Davies' data points.
The difference is due to the fact that the interacting level in the
$\mbox{NV}{}^{0}$ case has $A_{2}$ symmetry.

\begin{table}[hbtp!]
\begin{centering}
\caption{Stress parameters of the $\mbox{NV}^{-}$ and $\mbox{NV}{}^{0}$ ZPLs
in $\mbox{cm}^{-1}/\mbox{GPa}$. \label{tab:stressparameters}}

\par\end{centering}

\centering{} \begin{tabular}{cccc}
\hline 
 & $\mbox{NV}^{-}$ $1042.6\,\mbox{nm}$  & $\mbox{NV}^{-}$ $637\,\mbox{nm}$  & $\mbox{NV}{}^{0}$ $575\,\mbox{nm}$ \tabularnewline
\hline 
$A1$  & 3.9  & 12.3  & 8.5 \tabularnewline
$A2$  & -3.05  & -31.5  & -28.6 \tabularnewline
$B$  & 9.85  & 7.96  & 12.5 \tabularnewline
$C$  & 5.59  & 14.26  & 14.1 \tabularnewline
\hline
\end{tabular}
\end{table}

The extra level at $\sim115\,\mbox{cm}^{-1}$ for both the $\mbox{NV}^{-}$
and $\mbox{NV}{}^{0}$ centres is attributed to the occurrence of
a dynamic Jahn-Teller effect. The first vibronic state has $E\times E$
symmetry, which is split by linear electron-vibration interaction
into $E+(A_{1}+A_{2})$ with the $E$ displaced upward and $(A_{1}+A_{2})$
downward. The $A_{1}$ and $A_{2}$ degeneracy is lifted by quadratic
interaction where the order depends on sign: $A_{1}$ lower for $\mbox{NV}^{-}$
and $A_{2}$ lower for $\mbox{NV}{}^{0}$. The interaction can involve
a distribution of vibrations but when displaced down in energy to
where there is a low density of vibrational states the feature becomes
sharp. The $^{1}\! A_{1}$ transition from the emitting $A_{1}$ level
to the $A_{1}$ vibronic state is allowed by group theory but has
no oscillator strength and requires the mixing before the transition
can be observed. The energy gap to the next higher vibronic state,
$A_{2}$, is too large to have sufficient mixing to be observed. The
emission sideband at $344\,\mbox{cm}^{-1}$ can be attributed to the
upper $E$ vibronic state shifted slightly up in energy. The size
of the quadratic splitting is uncertain but indications are that there
is pseudo-localised mode at $250\,\mbox{cm}^{-1}$ and a Jahn-Teller
energy of order of $200\,\mbox{cm}^{-1}$.

\section{Electronic structure}

\begin{figure}
\includegraphics{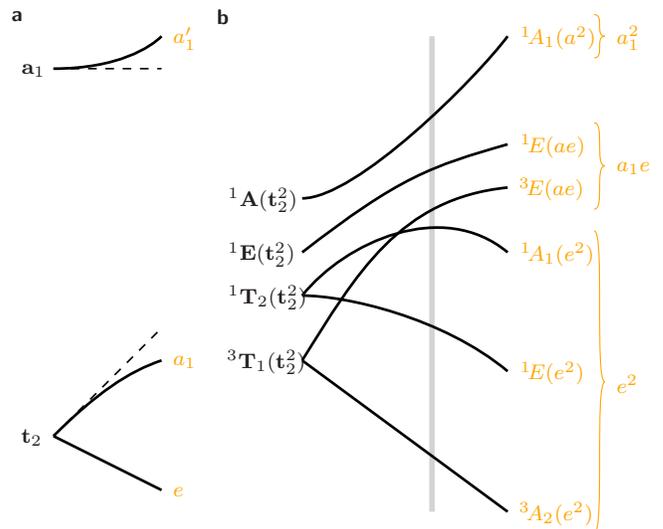}

\caption{\textbf{Trigonal from tetrahedral.} \textsf{\textbf{a}}, Single-hole
orbital energy scheme as $t_{d}$ tetrahedral symmetry is distorted
towards the $C_{3v}$ trigonal limit. Tetrahedral symmetry labels
are in bold, and the orange labels indicate the trigonal limit. \textsf{\textbf{b}},
Two-hole orbital energy scheme for the same distortion, where the
shaded grey region indicates approximately the $\mbox{NV}^{-}$ situation.
\label{fig:Trigonal-from-tetrahedral.}}

\end{figure}

A full treatment of the electronic structure of the $\mbox{NV}^{-}$centre
has been presented in a recent publication \cite{doherty2010_negatively},
whereas here we obtain insight by considering the centre as a trigonally
distorted vacancy centre of six electrons in order to identify the
correlations between the centre's different electronic states. In
the trigonally distorted vacancy picture, the molecular orbitals (MOs)
of the centre are associated with the linear combinations of the dangling
bonds in $T_{d}$ point group symmetry, such that they transform as
$A_{1}$ and $T_{2}$ and can be identified as $\mathbf{a}_{1e}$
and $\mathbf{t}_{2e}$. The $\mathbf{a}_{1e}$ is the lower energy
MO, and so for six electrons, the lowest energy electronic configuration
is $\mathbf{a}_{1e}^{2}\mathbf{t}_{2e}^{4}$. This can be more conveniently
described in terms of a two hole system, $\mathbf{t}_{2}^{2}$. Lowering
the symmetry from $T_{d}$ to $C_{3v}$ lifts the $\mathbf{t}_{2}$
degeneracy to give a two-fold degenerate $\mathbf{t}_{x,y}=e_{x,y}$
MO and a non-degenerate $\mathbf{t}_{z}=a_{1}$ MO. The $e$ MO is
independent of the strength of the trigonal field whereas the $a_{1}$
MO depends on the strength of the trigonal field as it is mixed with
the higher energy $a_{1}^{\prime}$ MO of the same symmetry associated
with $\mathbf{a}_{1}$ (figure \ref{fig:Trigonal-from-tetrahedral.}a).
In the large trigonal field limit the two $A_{1}$ MOs ($a_{1}$ and
$a_{1}^{\prime}$) are related to the tetrahedral MOs as $a_{1}=\sqrt{3}/2\mathbf{a}_{1}+1/2\mathbf{t}_{z}$
and $a_{1}^{\prime}=1/2\mathbf{a}_{1}-\sqrt{3}/2\mathbf{t}_{z}$.

Adopting the two-hole picture, the electronic configuration states
generated by the different occupations of the hole MOs are depicted
in figure \ref{fig:Trigonal-from-tetrahedral.} alongside their energies
as functions of the hypothetical trigonal field. The energetic ordering
of the states in both tetrahedral and strong trigonal field limits
are taken to be consistent with Hund's rules. The $\mathbf{t}_{2}^{2}$
configuration gives rise to the configuration states $^{3}\!\mathbf{T}_{1}(\mathbf{t}_{2}^{2})$,
$^{1}\!\mathbf{T}_{2}(\mathbf{t}_{2}^{2})$, $^{1}\!\mathbf{E}(\mathbf{t}_{2}^{2})$,
and $^{1}\!\mathbf{A}_{1}(\mathbf{t}_{2}^{2})$. The $^{3}\mathbf{T}_{1}(\mathbf{t}_{2}^{2})$
state splits to give the lower $^{3}\! A_{2}(e^{2})$ and the upper
$^{3}\! E(ea_{1})$ states in the trigonal field. The ground state
$^{3}\! A_{2}(e^{2})$ is independent of the size of the trigonal
field, whereas the $^{3}\! E(ea_{1})$ wavefunction varies with $a_{1}$.
It is known that there is a large nitrogen hyperfine interaction \cite{fuchs2008_excited-state}
and so $a_{1}$ has significant component of the nitrogen dangling
bond. The $^{1}\mathbf{T}_{2}(\mathbf{t}_{2}^{2})$ splits with the
$^{1}\! E$ decreasing in energy and the $^{1}\! A_{1}$ increasing
in energy, but both interact with higher energy states of the same
symmetry. In general, the states are admixtures $^{1}\! A_{1}={}^{1}\! A(e^{2})+\kappa^{\prime}A(a_{1}^{2})$,
$^{1}\! A_{1}^{\prime}={}^{1}\! A(a_{1}^{2})-\kappa^{\prime}\,^{1}\! A(e^{2})$
and $^{1}\! E={}^{1}\! E(e^{2})+\kappa E(ea_{1})$, $^{1}\! E={}^{1}\! E(ea_{1})-\kappa\,^{1}\! E(e^{2})$.

The presence of the $^{1}\! A_{1}-{}^{1}\! E$ transition indicates
the situation is far from the large trigonal limit, where it would
have zero oscillator strength. In that limit the ZPL would not be
displaced by $A_{1}$ stress components, but the nonzero $A1$ and
$A2$ parameters indicate that it is in fact displaced. Furthermore,
the $^{1}\! E\approx$$^{1}\! E(e^{2}$) state would have exactly
equal contribution of $e_{x}$ and $e_{y}$ MOs in the large trigonal
field limit and would thus not split with $E$ stress components,
which is clearly inconsistent with the observed splitting ($B$ and
$C$ parameters). From the magnitude of the stress parameters of the
$^{1}\! A_{1}-^{1}\! E$ transition, particularly in comparison to
the $^{3}\! A_{2}-$$^{3}\! E$ transition, an estimate can be made
for $\kappa$ to be of the order of 0.3. In the large trigonal field
limit $\kappa$ is zero, and -iIn $T_{d}$ symmetry the $^{1}\! E$
state can be re-expressed as $^{1}\! E=$$\sqrt{2/3}$$^{1}\! E(e^{2})+\sqrt{1/3}$$^{1}\! E(ea_{1})$
corresponding to a $\kappa$ value of 0.7. Hence, the observed $\mbox{NV}^{-}$
situation with $\kappa$ = 0.3 is intermediary and is indicated by
the vertical line in figure \ref{fig:Trigonal-from-tetrahedral.}b.

\begin{figure}
\includegraphics{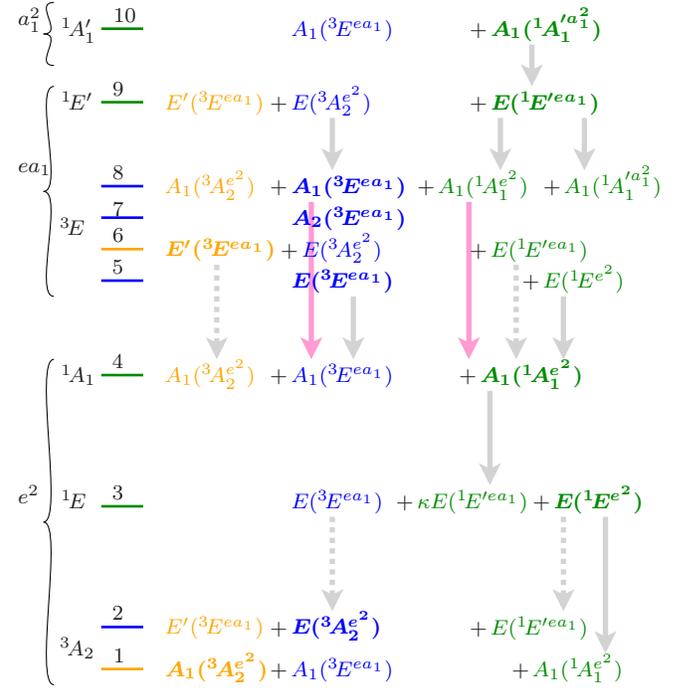}

\caption{\textbf{Mixing between levels.} The first column (orange) contains
$m_{s}=0$ components, the second column (blue) contains $m_{s}=\pm1$
components, and the final column (green) contains the $S=0$ singlets.
The primary two-hole configuration for each level is indicated in
bold, and the other terms in ordinary typeface are admixed components.
The coefficient $\kappa$ draws attention to the mixing in the lower
$^{1}\! E$ level. \label{fig:Mixing-between-levels.}}

\end{figure}

Spin-orbit interaction can be included in the above model of the configuration
states arising from the $\mathbf{t}_{2}^{2}$ configuration by considering
the variation of the transverse and axial spin-orbit parameters with
the trigonal field. In the $\mathbf{t}_{2}^{2}$ configuration, the
spin-orbit interaction is given by \begin{equation}
V_{so}=\Lambda_{x}S_{x}+\Lambda_{y}S_{y}+\Lambda_{z}S_{z}\end{equation}
 where $\Lambda_{x}$ and $\Lambda_{y}$ are the transverse spin-orbit
parameters, $\Lambda_{z}$ is the axial spin-orbit parameter, and
$S_{i}$ are electronic spin operators for $i=x,y,z$. In $T_{d}$
symmetry $\Lambda_{x}=\Lambda_{y}=\Lambda_{z}$, whereas in $C_{3v}$
the $\Lambda_{x}=\Lambda_{y}\neq\Lambda_{z}$. Note that a detailed
treatment of spin-orbit and spin-spin interactions was included in
the recent publication \cite{doherty2010_negatively}. The interactions
give the fine structure and this is shown in the energy level schematic
in figure \ref{fig:Mixing-between-levels.}. What is very important
is that the triplet states are split into states either with $m_{s}=0$
or with $m_{s}=\pm1$ spin projection. There is one minor mixing which
will be mentioned later. Optical transitions do not change spin projection
and so the centre will remain in one of the spin projections after
optical excitation. However, spin-orbit interaction will mix singlets
with either of the triplet spin projections and a change of spin projection
can result through intersystem crossings. An intersystem crossing
will occur non-radiatively through interactions with vibrations, but
the electron-vibration interaction can not by itself change spin projection.
For this reason the energy levels of the $\mbox{NV}^{-}$ centre are
presented in figure \ref{fig:Mixing-between-levels.} with the admixed
components arranged in three columns for $m_{s}=0$ , $m_{s}=\pm1$
and $S=0$. The figure enables the allowed intersystem decay to be
readily determined.

The most common situation is for the non-radiative decay to occur
through $A_{1}$ vibrations. For this, the two states involved in
the intersystem crossing have not only to be of the same irreducible
representation but have to have an admixture of a common state. There
is a further restriction in that the vibrational decay will be weak
for large energy gaps requiring many vibrations and so only adjacent
states are considered. By inspecting figure \ref{fig:Mixing-between-levels.}
it can be seen that the $A_{1}$ spin-orbit component of $^{3}\! E(ea_{1})$
and the $^{1}\! A_{1}$ singlet level are intermixed and, therefore,
allow decay via $A_{1}$ vibrations. For the same reason there can
be decay from $^{1}\! E^{\prime}$ to $^{3}\! E(ea_{1})$ but the
singlet is unlikely to be populated if above the triplet. There can
be radiative decay to the $^{1}\! E$ but there is no allowed decay
path from the lowest singlet. The minimal decay in the upper levels
and none returning the system to the ground state makes it clear that
optically induced spin polarisation can not occur through spin-orbit
interaction and $A_{1}$ vibrations alone.

\section{Electron-vibration interaction induced intersystem crossing}

The factor not considered in the above is electron-vibration interaction.
It is potentially a significant factor, as it can enable alternative
non-radiative decay between states as will be discussed in this section,
and dynamic effects within a state as discussed in the next section.
A tetrahedron has symmetry adapted displacements with $A_{1}$, $E$
and $T_{2}$ symmetry giving 2$A_{1}$ and 2$E$ distortions in trigonal
symmetry. The linear electron-vibration interaction associated with
such symmetry adapted vibrations can be written in $T_{d}$ as \begin{eqnarray}
H_{e-v} & = & V_{A_{1}}Q_{A_{1}}+V_{E_{x}}Q_{E_{x}}+V_{E_{y}}Q_{E_{y}}+V_{T_{2}x}Q_{T_{2}x}\notag\\
 & + & V_{T_{2}y}Q_{T_{2}y}+V_{T_{2}z}Q_{T_{2}z}\end{eqnarray}
 and in $C_{3v}$ \begin{eqnarray}
H_{e-v} & = & V_{A_{1}}Q_{A_{1}}+V_{E_{x}}Q_{E_{x}}+V_{E_{y}}Q_{E_{y}}+V_{E_{x}^{t}}Q_{E_{x}^{t}}\notag\\
 & + & V_{E_{y}^{t}}Q_{E_{y}^{t}}+V_{A_{1}^{t}}Q_{A_{1}^{t}}\end{eqnarray}

The additional decay channels introduced by electron-vibration interaction
are those allowed by $E$-vibrations and can be determined from figure
\ref{fig:Mixing-between-levels.}. These involve transitions where
there is a change of the one electron orbits from $a_{1}$ to $e$
or between the components of $e$ ($e_{x}$ to $e_{y}$). In the case
of the lowest transition between the $^{1}\! E$ and the $^{3}\! A_{2}$
ground state, decay is allowed to both of the ground state components
as indicated in figure \ref{fig:Mixing-between-levels.}. The decay
to the $E$ ground state component ($m_{s}=\pm1$) involves a change
between $a_{1}$ and $e$ orbits, whereas the decay to the $A_{1}$
component ($m_{s}=0$ ) involves a change of the $e$ orbits. As all
the spin-orbit mixings are for well separated states they will be
of similar magnitude and, therefore, the relative decay rates will
depend on these different orbit changes. The electron orbits for both
initial and final states lie largely on the three adjacent carbons
and it is likely that the $E$-vibration involved in the intersystem
crossing will be that involving the in-plane vibration of these atoms.
The reduced matrix elements of electron-vibration interaction associated
with this $E$ vibration are related to the reduced matrix elements
for tetrahedral symmetry, $\rdm{t_{x,y}}{V_{E}}{t_{x,y}}$, $\rdm{t_{z}}{V_{E}}{t_{x,y}}$
and $\rdm{a_{1}}{V_{E}}{t_{x,y}}$. The first two are equal whereas
the third is zero as $T_{2}\otimes E$ does not contain $A_{1}$.
In the large trigonal field limit, where $a_{1}$ is a linear combination
of $\mathbf{a}_{1}$ and $\mathbf{t}_{z}$ as given earlier, it is
found that the reduced matrix element $\rdm{a_{1}}{V_{E}}{e}=\sqrt{1/4}\rdm{e}{V_{E}}{e}$.
It has been shown above that $\mbox{NV}^{-}$ is not in the extreme
trigonal field limit, but $\rdm{a_{1}}{V_{E}}{e}$ will be still smaller
than $\rdm{e}{V_{E}}{e}$ (estimate $1/\sqrt{3}$). From this it can
be deduced that the reduced matrix elements associated with the transition
to the $A_{1}$ component ($m_{s}=0$) will be stronger than those
to the $E$ component $m_{s}=\pm1$ of the ground state by a factor
of 3/2. Thus, the decay will favour the population of the $m_{s}=0$
level of the ground state spin triplet.

The upper triplet-singlet $^{3}\! E$ - $^{1}\! A_{1}$ inter-system
crossing is also of great importance. As discussed above, relaxation
from the $A_{1}$ spin-orbit component of $^{3}\! E$ can decay to
the singlet via $A_{1}$-symmetry vibrations, but with electron-vibration
interaction there are four additional decay channels from the $E$
states allowed by $E$-vibrations. As the states involve $ea_{1}$
and $e^{2}$ configurations there is not the same justification for
considering only one of the $E$-symmetry vibrations. The four decay
channels are, however, not all equal as there will be a large variation
in the magnitude of spin-orbit mixing. In particular two of these
are large owing to the $^{1}\! A_{1}$ and $^{1}\! E'$ singlets undoubtedly
lying adjacent to the $^{3}\! E$ excited state triplet. The exact
positions of the singlets are not yet known and so the relative strength
of the two $E$-vibration induced relaxation and their strength relative
to the $A_{1}$-vibration induced relaxation cannot be estimated theoretically.
However, the relative intersystem crossing rates can be obtained experimentally
by comparing the emission responses upon switching on a pumping field
for the cases where the system is unpolarised, polarised by previous
optical excitation or inverted polarisation utilising a microwave
$\pi$ pulse. Such measurements have been given for ensembles \cite{manson2006_nitrogen-vacancy}
and single sites \cite{steiner2010_universal,robledo2010_spin}. The
crossing rate for $m_{s}=\pm1$ states are estimated to be factor
of six times faster than that for $m_{s}=0$ states \cite{robledo2010_spin}.
This requires the decay involving $A_{1}$ vibrations to be slightly
more than a factor of two faster than those involving $E$ vibrations.

The electron-vibration interaction is also important for the decay
between the singlets, $^{1}\! A_{1}$ - $^{1}\! E$. It does not require
spin-orbit mixing and consequently can be a strong transition and
is known to dominate over the radiative transition between these levels
\cite{rogers2008_infrared}. In this and other cases we have only
discussed the decay in terms of the linear process implying only one
$E$-vibration. However, this is the enabling process and other $A_{1}$
vibrations or pairs of $E$ vibrations will be involved in compensating
for the energy gap.

\section{Electron-vibration interaction within electronic states}

A degenerate vibration interacting with an orbitally degenerate electronic
state can give rise to a Jahn-Teller or dynamic Jahn-Teller (DJT)
effect and the uniaxial stress measurements showed that there was
just such an effect in the lowest singlet state $^{1}\! E$. In the
previous section it has been argued that there is spontaneous decay
to the ground state via $E$ vibrations. This does not require vibrations
to be present in the lattice but there is the possibility of an additional
process at higher temperatures when the lowest $A_{1}$ vibronic level
is populated. From this $A_{1}$ state there can be decay to the ground
state with the annihilation of the $E$ vibration combined with additional
$A_{1}$ vibrations or pairs of $E$-vibrations to make up for the
energy mismatch (anticipated to be of the order of $5000\mbox{\,}\mbox{cm}^{-1}$).
The additional decay path will result in a temperature dependent $^{1}\! E$
lifetime and this has been observed in both absorption \cite{acosta2010_optical}
and emission recovery \cite{robledo2010_spin}. The two papers both
reported a temperature dependence consistent with populating a level
at $128\,\mbox{cm}^{-1}$ ($16\,\mbox{meV}$). The variance from the
$115\,\mbox{cm}^{-1}$ observed here will be due to their value corresponding
to a distribution average and this one to a peak.

The presence of DJT is already known to occur in the excited $^{3}\! E$
state \cite{fu2009_observation}. The electron-vibration interaction
results in a change of the polarisation of the sideband from that
of the zero-phonon transition \cite{fu2009_observation} and the interaction
with the low frequency distribution of $E$-vibrations leads to a
$T^{5}$ broadening of the zero-phonon line rather than a more normal
$T{}^{7}$ dependence \cite{fu2009_observation}. There will be some
quenching of the spin-orbit splitting, but the degree is unclear as
the magnitude of the intrinsic spin-orbit is not known.

Another important consequence of the DJT effect in the excited $^{3}\! E$
state is that for a fixed spin it mixes the orbital components (no
$m_{s}=0$ and $m_{s}=\pm1$ mixing). The $E_{x}$ component of the
electron-vibration interaction mixes the $A_{1}$ spin-orbit level
with the $E_{x}$ level and enables population in the $E_{x}$ state
to decay via $A_{1}$ vibrations, and also $A_{1}$ population to
decay via $E$ vibrations. There is a similar situation for the $E_{y}$
component of the interaction involving the $E_{y}$ state. Note also
that population in the $A_{2}$ state can decay because of such mixing.
The consequence is that even at zero temperature the DJT will enhance
the overall decay rates.

\begin{figure}
\includegraphics{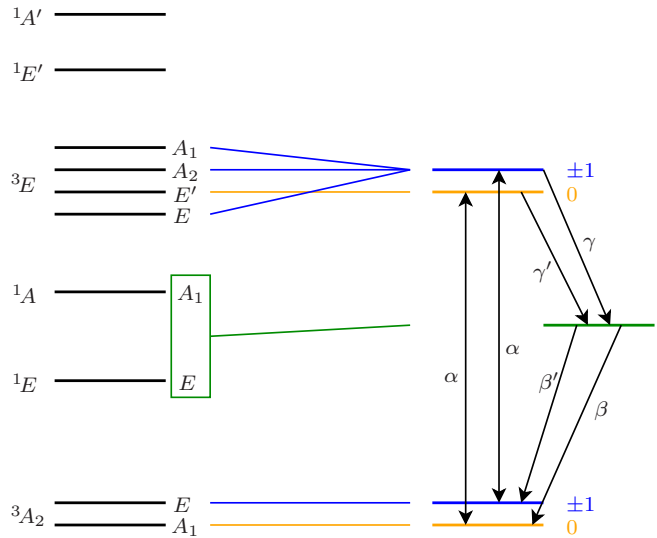}

\caption{\textbf{Rate equation model.} At room temperature the $\mbox{NV}^{-}$
system may be approximated as a 5-level model. \label{fig:Rate-equation-model.} }

\end{figure}

As temperature is increased there can be a real transfer of population
between the $^{3}\! E$ levels through a two $E$ vibration process.
The rate can become higher than the separation of the levels. This
quenches orbital angular momentum and the result is that the excited
$^{3}\! E$ state behaves as an effective orbital singlet (figure
\ref{fig:Rate-equation-model.}). There are three spin projections
in the ground and three in the excited state but in each case the
$m_{s}=\pm1$ states can be treated as one. The decay rate between
the singlets is fast and so they can be treated as one level. Thus
the dynamics of the centre can be modeled using 5 levels, and this
is frequently the electronic model used in discussing room temperature
observations \cite{manson2006_nitrogen-vacancy,rogers2009_time-averaging,robledo2010_spin}.

\section{Rate equations for spin polarisation}

The 5-level system is indicated in figure \ref{fig:Rate-equation-model.}.
The effective triplet-singlet intersystem crossing rates are the sum
for the various $^{3}\! E$ spin-orbit components, and are denoted
by $\gamma$ and $\gamma'$ for transfer from $m_{s}=\pm1$ and $m_{s}=0$
respectively. The lower crossings are denoted by $\beta$ and $\beta'.$
Anticipating spin polarisation the dashed parameters are expected
to be slower than the non-dashed parameters. The optical transition
strength is taken as $\alpha$. The model does not include a small
mixing between the $E$ and $E'$ spin-orbit components of $^{3}\! E$
state by spin-spin interaction. It is less than 1 percent and can
be neglected for the case where the spin polarisation is not greater
than 90 per cent. Assuming that the spin lattice relaxation is indefinitely
long then the degree of spin polarisation is independent of the strength
of the optical field. The population ratios in the excited state and
ground states during continuous optical pumping are given by \begin{eqnarray}
\frac{P_{\pm1}^{e}}{P_{0}^{e}} & = & \frac{\gamma'}{\gamma}\frac{\beta'}{\beta}\\
\frac{P_{\pm1}^{g}}{P_{0}^{g}} & = & \frac{\alpha+\gamma}{\alpha+\gamma'}\frac{\gamma'}{\gamma}\frac{\beta'}{\beta}\end{eqnarray}

Due to the faster intersystem crossing for $m_{s}=\pm1$ states, there
is always a smaller ratio (higher spin polarisation) in the excited
state. $(\alpha+\gamma)/(\alpha+\gamma')$ is the ratio of the lifetimes
$12.0\,\mbox{ns}/7.8\,\mbox{ns}=1.54$ \cite{batalov2008_temporal}.
Different values will be obtained when the optical pumping is stopped.
The polarisation will immediately reduce but by what amount depends
on the $\beta'/\beta$ ratio. Should high pump powers be used, so
that prior to turning off the pump the majority of the population
is in the singlet states, then after relaxation the polarisation will
be determined largely by the $\beta'/\beta$ ratio. Our above estimates
suggest that this ratio is not very supportive of spin polarisation
and it would be better to use low intensities. However, it does also
indicate an opportunity to measure this ratio by examining spin polarisation
as a function of the strength of the preparation pulse.

The intersystem crossing ratio to the ground state was estimated as
$\beta'/\beta=2/3$ and the upper crossing given earlier gives $\gamma'/\gamma=1/6$.
Consequently $(\gamma'/\gamma)(\beta'/\beta)=2/18=0.11$ and for a
lifetime ratio of $1/6$ corresponds to a ground state ratio of 0.18.
This indicates a ground state spin polarisation of 82 percent in the
$m_{s}=0$ state. This makes plausible comparison with experimental
values of spin polarisation as they are also of the order of 80 per
cent \cite{robledo2010_spin,neumann2010_quantum,fuchs2010_excited-state,harrison2006_measurement}.
Higher spin polarisation values would mean that the ratio $\beta'/\beta$
is much smaller than our estimate. This would be the case should the
reduced matrix element $\rdm{a_{1}}{V_{E}}{e}$ for the participating
$E$ vibration be very small but better estimates will require much
more advanced theoretical modeling. However, it would be better to
first establish the $\beta'/\beta$ ratio experimentally.

\section{Discussion and conclusions}

It should be recognised that due to the nature of spin-orbit interaction,
intersystem crossings will always be spin selective and, hence, one
can expect optical induced spin polarisation for all systems with
$S\geq1$ ground states. Whether this is observed depends on other
factors. Obviously the centre needs to be stable and spin lattice
relaxation times reasonably long. The important aspect highlighted
by the present work is the rate and sign of the intersystem crossing.
There will be two crossings, and for significant spin polarisation
it is beneficial to have a high branching ratio between the alternative
spin projections for the two crossings to favour the same spin. The
rates will be faster if the levels are close and this is more readily
achieved if there are two intermediate levels (such as the two singlets
in the case of $\mbox{NV}^{-}$). An aspect not treated above is that
optical spin readout requires the upper inter-system rate to be comparable
to the optical rate, and this is probably only achievable if the levels
are close to one phonon energy.

Clearly NV satisfies all of the above, but it should be realised that
this is only achieved as a consequence of electron-vibration interaction.
The upper intersystem crossing is between two close levels and although
a reasonable rate maybe achievable without the vibration interaction
the rate is certainly enhanced with vibrations. The vibration interaction
enables additional decay paths and also improves the situation with
mixing within the manifold through the DJT effect. The situation is
more important for the lower intersystem crossing where without the
electron-vibration interaction the lower singlet level would be metastable
and population would simply be transferred from triplet to singlet.
This is probably what happens in several other colour centres in diamond
where the alternate spin state is sufficiently long lived, for example,
to allow CW EPR measurements \cite{felton2008_electron}. The interaction
involves low frequency vibration and the specifics of the vibration
are important as it can influence the rates for the different spins.
It will not be a simple procedure to identify another `NV situation'.
It may be easy to have equivalent energy structures even in similar
lattices, but it is hard to anticipate how to obtain appropriate strength
of electron-vibration interaction and appropriate local vibrations.

The main point of the paper has been to show that optically induced
spin polarisation can only be explained when the effect of electron-vibration
interaction is included alongside the electronic spin-orbit interaction
in the dynamics of the centre. By using the current electronic model
and including simple considerations of electron-vibration interaction
it is shown that the estimated spin polarisation is in agreement with
experiment.
\begin{acknowledgments}
This work was supported by the Australian Research Council under the
Discovery Project scheme DP0986635 and DPxxxxxxx. 
\end{acknowledgments}

\end{document}